\documentclass{jkas}

\def\year{2014} 
\def\month{March} 
\def\volume{00} 
\def\issue{0} 
\def\beginpage{1} 
\def\endpage{99} 
\def\received{---} 
\def\accepted{---} 

\date{Received \received ; accepted \accepted}

\title{
    Microlensing by Kuiper, Oort, and Free-Floating Planets
}

\author[1,2,3]{Andrew Gould}

\affil[1]{Max-Planck-Institute for Astronomy, K\"onigstuhl 17, 69117 Heidelberg, Germany; \email{gould@astronomy.ohio-state.edu }}
\affil[2]{Korea Astronomy and Space Science Institute, Daejon 305-348, Republic of Korea}
\affil[3]{Department of Astronomy Ohio State University, 140 W.\ 18th Ave., Columbus, OH 43210, USA; \email{weizhu@astronomy.ohio-state.edu }}

\def\corrauthor{
Andrew Gould
}

\def\runningauthor{
Andrew Gould 
}

\def\runningtitle{
    Kuiper, Oort, and Free-Floating Planets
}

\def\keywords{
astrometry -- gravitational microlensing -- planets -- stars
}

\def\abstracttext{
Microlensing is generally thought to probe planetary systems only
out to a few Einstein radii.  Microlensing events generated by
bound planets beyond about 10 Einstein radii generally do not
yield any trace of their hosts, and so would be classified as 
free floating planets (FFPs).  I show that it is already possible,
using adaptive optics (AO), to constrain the presence of potential
hosts to FFP candidates  at separations comparable to the Oort Cloud.
With next-generation telescopes, planets at Kuiper-Belt separations 
can be probed.  Next generation telescopes will also permit
routine vetting for all FFP candidates, simply by obtaining
second epochs 4--8 years after the event.
At present, the search for such hosts is restricted to within the
``confusion limit'' of $\theta_\confus\sim 0.25^{\prime\prime}$, but
future {\it WFIRST} observations will allow one to probe beyond this
confusion limit as well.
}

\newcommand{\bdv}[1]{\mbox{\boldmath$#1$}}

\def\au{{\rm AU}}

\def\kms{{\rm km}\,{\rm s}^{-1}}
\def\masyr{{\rm mas}\,{\rm yr}^{-1}}

\def\kpc{{\rm kpc}}

\def\mas{{\rm mas}}
\def\lim{{\rm lim}}

\def\pc{{\rm pc}}

\def\max{{\rm max}}
\def\snow{{\rm snow}}

\def\rel{{\rm rel}}

\def\hel{{\rm hel}}
\def\psf{{\rm psf}}
\def\geo{{\rm geo}}
\def\confus{{\rm confus}}

\def\e{{\rm E}}
\def\bpi{{\bdv\pi}}
\def\bmu{{\bdv\mu}}

\def\apj{{ApJ}}
\def\apjl{{ApJL}}

\def\kms{{\rm km~s^{-1}}}
\def\au{{\rm AU}}

\begin{document}
\jkashead 


\section{{Introduction}
\label{sec:intro}}

\citet{sumi11} have reported evidence of a vast population of
free-floating planets (FFP) based on an excess of short timescale 
($t_\e\sim 1\,$day) microlensing events seen in observations toward
the Galactic bulge by the Microlensing Observations in Astrophysics
(MOA) collaboration.  As the authors recognize, these planets
are not necessarily unbound: they may simply be so far from their hosts
that the host leaves no trace on the microlensing event.
\citet{sumi11} searched for two such effects: (1) ``bumps'' (perhaps
of very low amplitude) due to microlensing by the host long before
or after the short ``FFP'' event; (2) distortions
in the short event (relative to a point lens event) due to shear
from the host.  For each FFP candidate, they quantified the limits that could
be put on hosts due to the absence of both effects.

Generically, one expects the hosts to be detectable only for
planets at projected separations of up to about 5 host Einstein 
radii.  Of course, any particular host might be detected at
much larger separation provided that the source trajectory is
sufficiently closely aligned with the planet-star separation axis.
For example \citet{ob08092} and \citet{mb13605} detected two
such examples, which they characterized as ``Uranus'' and ``Neptune''
analogs, respectively.  However, for a large fraction of such
ice-giant analogs, the host would have left no trace.  And this
would be even more true of more distant planets.

Planets that leave no trace may turn out to be in Kuiper-Belt
like orbits at $15\lesssim a/r_\snow\lesssim 150$, or in Oort-Cloud
like orbits $150\lesssim a/r_\snow\lesssim 10^4$.  Here 
\begin{equation}
r_\snow\sim 2.7\,\au {M\over M_\odot}
\label{eqn:snow}
\end{equation}
is the snow line, assumed to scale linearly with stellar mass.
The snow line can be related to the Einstein radius
$r_\e = D_L\theta_\e$ by
\begin{equation}
{r_\snow\over r_\e} = 2.7{\pi_L(M/M_\odot)\over(\kappa M\pi_\rel)^{1/2}}
=0.95
\biggl({M\over M_\odot}\biggr)^{1/2}
\biggl({D_L D_{LS}/ D_S \over \kpc}\biggr)^{-1/2}.
\label{eqn:snowrat}
\end{equation}
Here, $\kappa = 4 G/c^2\au=8.14\,\mas\,M_\odot^{-1}$, $\pi_L=\au/D_L$ is the
lens parallax, $\pi_\rel = \pi_L - \pi_S$ is the lens-source relative
parallax, and $D_{LS}=D_S - D_L$.  Thus, for typical lens masses
$(M\sim 0.3\,M_\odot)$ and distances $D_L D_{LS}/D_S \sim 1\,\kpc$,
we have $r_\snow \simeq 0.5\,r_\e$.  Therefore, Kuiper planets are
at least 7.5 Einstein radii from their hosts, and thus it is very unlikely
that their hosts will induce microlensing signatures on the ``FFP'' event.

\section{{Confusion Limit of Microlensing}
\label{sec:confusion}}

On very general grounds, the optical depth to microlensing is
of order $\tau\sim {\cal O}(v^2/c^2)$ where $v$ is the characteristic
velocity of the system.  For the Milky Way, this predicts $\tau\sim 10^{-6}$
which agrees with observations.  That is, the
probability that some lens will lie within $n$ Einstein radii
of a given location is $p=n^2\tau$.  Or, in other words, we expect
of order one such lens to lie withing $\theta_\confus = \tau^{-1/2}\theta_\e
\gtrsim 10^3\theta_\e$ of any given point.  Since typically
$\theta_\e\sim 0.25\,\mas$,
this corresponds to $\theta_\confus\sim 0.25^{\prime\prime}$.

In particular, for genuine FFPs, we expect the nearest star
(other than the source) to be separated by 
of order $\Delta\theta\sim \theta_\confus$.
The only exception would be a companion to the source.  Of course,
by chance, there could be unassociated stars that lie closer than
this.  However, the prior probability for, .e.g., an unassociated
star at $\Delta\theta =0.1\,\theta_\confus\sim 25\,\mas$ would be just 1\%.
While not much can be concluded from the detection of such a nearby star in any
particular case, even a few such detections in a sample of a few dozen
``FFPs'', would be evidence for a population of well-separated
planets.

Such closely separated stars cannot be resolved even in excellent-seeing
ground-based survey data.  Therefore, special efforts and/or new
space-based surveys are required to detect them.

\section{{Oort Planets from AO on Existing Telescope}
\label{sec:oort}}

\citet{Batista:2015} and \citet{Bennett:2015} have detected a lens
star that had separated from the source by $\sim 60\,\mas$
due to $\sim 8\,$yr of relative proper motion using Keck adaptive optics
(AO) and {\it Hubble Space Telescope (HST)} observations, respectively.  
In that case, the
lens and source were of comparable brightness.  Hence, it is plausible
that one could detect very faint hosts of FFPs at $\Delta\theta\sim 100\,\mas$.,
i.e., roughly $400\,\theta_\e$ or $800\,r_\snow$.

If such a star were detected, it would not automatically imply
that it was the host
to the ``FFP''.  The star could be a companion to the source, a companion
to the true host, which was itself much closer to the planet, or a random interloper.
The first of these possibilities is easily tested by taking a second
epoch of observations a few years later.  If the detected star
is a companion to the source, the two will have the same proper motion.

Then, assuming it passes this test, there are only three possibilities:
host, companion to the host, or random interloper.  Either of the
first two possibilities would imply that this is not an FFP.
As mentioned
above, the last would occur with probability 
$p\sim 16\%(\Delta/\theta/100\,\mas)^2$.  Hence, by searching for
such hosts of a few dozen FFP candidates, one could place significant
limits on such hosts (or confirm their existence).

However, even if the result of this statistical analysis were that
a significant fraction of putative FFPs had distant companions, this
still would not distinguish Oort planets from Kuiper planets
whose hosts happen to have substantially more distant companions.

\section{{Kuiper Planets from AO on Next Generation Telescopes}
\label{sec:kuiper}}

Next generation telescopes such as the Giant Magellan Telescope (GMT),
Thirty Meter Telescope (TMT) and Extremely Large Telescope (ELT)
will have diameters 2.5 to 4 times larger than Keck and will operate
effectively in $J$ band (compared to the $K$ band observations of
\citealt{Batista:2015}), and so will be able to search for hosts
that are closer by a factor 5--8.  To be concrete, I adopt a factor 6,
i.e., $\Delta\theta\sim 17\,\mas$ or about $70\,\theta_\e$.  This
is within the Kuiper range of planetary orbits that I defined above.  
The chance of
random interlopers is reduced by a factor $6^2=36$ at this limit,
i.e., to of order $p\sim 0.5\,\%$.  While, this is still not low enough
to absolutely rule out this possibility in any individual case,
it would enable very strong statistical statements with even
three detections out of a few dozen searches.  
Moreover, the possibility that the detected
star is a companion to the host rather than the host itself
will be far more restricted for the close-in detections enabled
by next-gen telescopes.  That is, an FFP candidate would have
to be a least a few $\theta_\e$ from a putative host to avoid 
the two detectable effects mentioned above (bump from host and
distortion of planetary event due to shear from host).  And
hierarchical stability typically requires factor $\gtrsim 3$ ratio
in semi-major axes.  These considerations cannot be used to
rule out such companions, partly because at $\Delta\theta\sim 70\,\theta_\e$,
there is still easily enough room for such hierarchies.  Moreover,
due to projection effects, it is possible for companions that
are well separated in three space to have similar projected position.
Nevertheless, these scenarios would be far more restricted than
what is achievable with present-day telescopes.

Such cases could be further constrained by followup observations
taken a few years after the event when the source and lens had
separated.  If the true host of the planet is much closer than
the star that was separately resolved at the time of the event,
then three stars will be observed in the late time observations,
including the source star, the host, and its more distant companion,
with the latter two having the same proper motion.

\section{{AO Imaging of All FFP Candidates}
\label{sec:ao}}

Indeed, all FFP candidates can be directly tested for distant
companions ``simply'' by obtaining two high resolution images: one at
the time of the event and one at a sufficiently later time that
the source and lens can be separately resolved.  If the planet
has a luminous host (i.e., not a brown dwarf, neutron star or black hole),
this will appear next to the source at the second epoch.

I have placed ``simply'' in quotation marks because with present
technology, the second epoch must wait about 25 years to achieve
$100\,\mas$ separation, even assuming typical proper motions
of $\mu=4\,\masyr$.  To be relatively certain that the
failure to observe the host was not just due to abnormally low
proper motion, one should really wait 50 years. 

However, with next generation telescopes, these numbers will each 
come down by a factor 6.  Hence, one could obtain a good statistical
understanding after just 4 years, and an excellent one after 8 years.

\section{{Probing Kuiper Planets with {\it WFIRST}}
\label{sec:wfirst}}

{\it WFIRST} is a planned NASA mission, currently in Phase A.
It will have a substantial microlensing component, probably consisting
of six 72-day campaigns centered on quadrature and covering $2.8\,\rm deg^2$
with a 15 min cadence \citep{Spergel:2015}.  True FFPs detected by
{\it WFIRST} should be point-lens events with blended light fractions
consistent with zero.  In principle, there could be blended
light from a companion to the source or from a random interloper.
However, as mentioned above the first possibility is easily vetted
by checking for common proper motion with the source over
several years using AO imaging.  The prior probability of the second
is, as before,  $p=(\Delta\theta/\theta_\confus)^2$.  For example, if the
blended light is $f_b =10\%$ of the source light, and the putative FFP event
is detected with total $(\rm S/N)=100$, then the offset between the blend
and the source (and so lens) can be measured with precision 
$\sigma(\Delta\theta)\sim \theta_\psf/(f_b({\rm S/N}))\rightarrow 10\,\mas$.
Here $\theta_\psf$ is the characteristic size of the point spread function.
Thus, these measurements could both reliably detect offsets and make 
strong statistical statements against interlopers.

\section{{Beyond the Confusion Limit}
\label{sec:beyond}}

The confusion limit of $\theta_\confus = 0.25^{\prime\prime}$ 
corresponds to about 2000 AU, which is well inside the zone
containing the Oort Cloud comets that come by the Sun after being
perturbed by random stars.  Moreover, it would be difficult to
reliably identify a star even a factor few closer than this
limit as the host of a putative FFP based on statistical arguments
alone.  Is it possible to probe out to -- or beyond -- this
confusion limit?

\citet{zhugould} have analyzed the potential for measuring
microlens parallaxes of FFPs by simultaneously observing the
{\it WFIRST} fields from a network of ground-based observatories.
See also \citet{han04} and \citet{yee13}.  \citet{zhugould}
showed that particularly for low
mass FFPs ($m_p\lesssim M_{\rm jup}$), measurements of the full 2-D 
microlens parallax $\bpi_\e$ are possible.  Here 
\begin{equation}
\bpi_\e = \pi_\e{\bmu_\rel\over\mu_\rel};
\qquad
\pi_\e = {\pi_\rel\over\theta_\e}
\label{eqn:bpie}
\end{equation}
\citep{gould92},
where $\bmu_\rel$ is the lens-source relative proper motion.

Each candidate host that is observed in AO images can be
vetted as follows.  First one can check whether the direction
of proper motion $\bmu_\rel$ is consistent with the direction of the 
$\bpi_\e$ parallax of the FFP.  Now, in fact, there is a slight
wrinkle here because $\bpi_\e$ is measured in the geocentric
frame whereas the relative proper motion of the putative host and
the source would be measured in the geocentric frame.  These differ
by
\begin{equation}
\delta\bmu_\rel=
\bmu_{\rel,\hel} - \bmu_{\rel,\geo} = {\pi_\rel\over\au}{\bf v}_{\oplus,\perp}
\label{eqn:bpiedif}
\end{equation}
where ${\bf v}_{\oplus,\perp}$ is the velocity of Earth projected on the
sky at the time of the event. However, first $v_{\oplus,\perp}$ is
typically small for {\it WFIRST} because the observations are 
centered at quadrature.  Second, $\pi_\rel$ of the putative host
can be estimated photometrically.  Hence, $\delta\bmu_\rel$ is both
small and partly calculable.  For example, if $v_{\oplus,\perp} =10\,\kms$
and one estimates $\pi_\rel$ with an error of 
$\sigma(\pi_\rel) = 0.03\,\mas$, then the error in the estimate of
this difference is
$|\sigma(\delta\mu_\rel|)\sim 0.06\,\masyr$, which is quite
small compared to typical proper motions of microlensing events,
$\mu_\rel\sim 4\,\masyr$.

Second, in some cases, one can estimate the magnitude of the proper
motion as well as the direction.  That is,
$\mu_\rel = \theta_\e t_\e = \pi_\rel t_\e/\pi_\e$.  For putative hosts
in the bulge, $\pi_\rel$ cannot be estimated precisely enough for
this equation to be useful.  However, for hosts in the disk,
$\pi_\rel$ can often be estimated photometrically to much better
than a factor 2.  Hence, for these, both the magnitude and
direction of $\bmu_\rel$ can be predicted well enough to
eliminate most random interlopers as hosts.

Finally, in a significant minority of cases for which $\bpi_\e$ can
be measured by combined ground-based and {\it WFIRST} observations, 
$\theta_\e$ can also be measured \citep{zhugould}.  For these,
the full proper motion of putative hosts can be predicted.  Moreover,
their photometrically
estimated relative parallax must be consistent with that of the
FFP, i.e., $\pi_\rel = \pi_\e\theta_\e$.

It is notable that very low mass FFPs, i.e., in the Mars--Earth range,
are the most likely to yield parallax measurements.  If the mechanisms
that drive objects into Oort-Cloud like orbits are similar to those
in the Solar System, then these are most feasibly applied to such
low mass objects.

\section{Dynamical Stability of Oort Planets}
\label{sec:stable}

If planets were launched into orbits similar to those of Oort-Cloud
comets when a given planetary system formed, would they remain bound
until today?

To address this I consider first the deflection due to the
closest single encounter with a passing star (to either the planet
or the host), which has impact parameter $b$ satisfying
$2\pi b^2 n v T\simeq 1$, where $n$ is the number density of
ambient stars, $v$ is their typical velocity relative to the host,
and $T$ is the age of the planetary system.  In this closest
impact, the host or planet will be deflected in velocity by
$\delta v= 2Gm/bv$ where $m$ is the mass of the passing star.  
I then set $(\delta v)^2 = Gm^\prime/a$ as the condition to ionize
the planet, where $m^\prime$ is the host mass and $a$ is the semimajor
axis.  For simplicity, I take $m^\prime = m$ and then find
\begin{equation}
a_\max = {v\over 8\pi G\rho T}
\label{eqn:ionize}
\end{equation}
where $\rho\equiv nm$ is the stellar density.  For disk lenses,
$\rho\sim 0.05\,M_\odot$, $v\sim 50\,\kms$, and $T=5\,$Gyr,
which implies $a_\max\sim 2\,\pc$.  This is consistent with
the survival of the Oort Cloud in our own Solar System.

For basically self-gravitating stellar systems like the
Galactic bulge, one may first note that $v^2/8\pi G\rho\simeq R^2$
where $R$ is the size of the system.  Hence
\begin{equation}
a_\max \rightarrow {R^2\over vT}\,\qquad \rm (self-gravitating).
\label{eqn:ionize2}
\end{equation}
Adopting $R\sim 1\,\kpc$,
$T=10\,$Gyr and $v=150\,\kms$ for the bulge, one finds
$a_\max \sim 0.7\,\pc$.  

A more precise calculation would take account of diffusive processes
from multiple sub-ionizing encounters.  However, it is clear from
this calculation that planets in Oort-like orbits are permitted
even in the Galactic bulge.

\section{{Conclusion}
\label{sec:conclude}}

Microlensing is generally thought to probe planetary systems
on scales of the Einstein radius, or perhaps a few Einstein radii.
I have shown that by combining microlensing observations of seemingly
isolated planets with high resolution imaging, one can probe planetary
systems out to the confusion limit 
$\Delta\theta\sim\theta_\confus\sim 0.25^{\prime\prime}$, and even beyond.


\acknowledgments
I thank Scott Gaudi for stimulating discussions.
I thank the Max Planck Institute for Astronomy for its hospitality. 
This work was supported by NSF grant AST-1516842.


\end{document}